\documentclass[aps,twocolumn,prb,floatfix]{revtex4-2}
\usepackage{epsfig}
\usepackage{dcolumn}
\usepackage{bm}
\usepackage{amsmath}
\usepackage{tikz}
\usepackage{graphicx, graphics, color}
\usepackage{natbib}

\usepackage[colorlinks=true,citecolor=blue]{hyperref}

\ifpdf
	\DeclareGraphicsExtensions{.pdf,.png,.jpg,.mps}
\else
	\DeclareGraphicsExtensions{.eps}
\fi

\newcommand{\be}{\begin{equation}}
\newcommand{\ee}{\end{equation}}
\newcommand{\beq}{\begin{eqnarray}}
\newcommand{\eeq}{\end{eqnarray}}

\begin{document}

\title{
{Signatures of the Correlated-Hopping Interaction in Non-Linear Transport through a Quantum Dot}}

\author{Ulrich Eckern}
\affiliation{Institute of Physics, University of Augsburg, 86135 Augsburg, Germany\\
ulrich.eckern@physik.uni-augsburg.de\\
https://orcid.org/0000-0001-8917-9083}

\author{Karol I.\ Wysoki\'{n}ski}
\affiliation{Institute of Physics, M.~Curie-Sk{\l}odowska University, pl.~M.~Curie-Sk{\l}odowskiej 1, 20-031 Lublin, Poland\\
karol.wysokinski@umcs.lublin.pl\\
https://orcid.org/0000-0002-5366-4455}


\begin{abstract}
In condensed matter systems with the Coulomb interaction playing an important role one expects, besides the on-site (local) Hubbard-type interaction, that also other (non-local) terms depending on the site occupancy, known as correlated or assisted hopping, exist. Even though such terms in quantum dots tunnel coupled to external electrodes may have quite an appreciable amplitude, the interpretation of experiments on these systems---usually in the linear response regime---seems not to require their presence. However, since the correlated-hopping term breaks the particle-hole symmetry of the standard Anderson model and modifies all transport characteristics of the system, the detailed knowledge of its influence on measurable characteristics, especially {\em in the non-linear regime}, is a prerequisite for its experimental detection. In this paper, we study the non-linear transport properties of junctions composed of a quantum dot tunnel coupled to external electrodes. We model the system by the single-impurity Anderson Hamiltonian with Hubbard on-site interaction and with a non-local correlated-hopping interaction. Using the previously found general expression for the spin-dependent transport and spectral Green functions as well as general formulae for charge and heat transport, we calculate relevant transport characteristics in the strongly non-linear regime. 
\end{abstract}

\maketitle

\section{Introduction}
In narrow-band condensed matter systems, the Coulomb interaction plays a significant role. The magnitude of the interaction is typically characterized by the single parameter $U$, which has the physical interpretation as a local repulsion between two electrons on the same site. In Wannier representation with states $|i\rangle$ centered at site $i$, its magnitude is given by the diagonal matrix element, $U=\langle ii|V|ii\rangle$, of the Coulomb potential $V(\mathbf{r})$. In specific situations, other matrix elements may also be important. In particular, this is true for the term that connects two neighboring sites and has the structure of density-dependent hopping, $K_{ij}=\langle ii|V|ij\rangle$, also known as correlated hopping. The latter describes situations where the hopping of a particle depends on the presence of other particles at the sites involved in the process. Note that for the model studied below, $i$ refers to a dot state, and $j$ to a state in the leads.

Some time ago, the role of correlated hopping was studied in the context of superconductivity~\cite{micnas1991}, where this kind of interaction alone was shown to lead to a superconducting instability. More recent studies considered the possibility of simulating such and more complicated density-dependent interactions in optical lattices~\cite{liberto2014}. On the condensed matter side, besides its importance for superconductivity~\cite{wysokinskimm2017}, the role of such interactions in establishing the ferromagnetic many-body state~\cite{westerhout2022}, or interacting higher-order topological insulators~\cite{montorsi2022}, has been studied.

 Furthermore, both $U$ and $K$ are expected to be important in quantum dots~\cite{guinea2003,borda2004,stauber2004,lin2007,tooski2014,gorski2019} and point contacts~\cite{cronenwett2002,meir2002}. Interestingly, despite very precise measurements of the conductance in quantum dot structures including the detailed scaling of the charge conductance with temperature in the Kondo regime~\cite{goldhaber-gordon1998}, the experiments could be described solely by using the standard Hubbard model with the $U$ term only. However, it is important to stress that both theory and experiment concentrated on the linear regime so far, while non-linear transport in the model with correlated hopping has gained comparatively little attention~\cite{eckern2021}. It is the purpose of the work to fill this gap and study the effect of correlated hopping on the conductance and the thermoelectric power in the strongly non-linear regime where specifically a large voltage bias and/or temperature difference is applied. To reach this goal, we use the non-equilibrium Green function (GF) approach. We model the system by the generalized Anderson Hamiltonian with the correlated-hopping term as discussed earlier~\cite{eckern2021}.

We find, in particular, a strong asymmetry in the conductance and the Seebeck coefficient corresponding to the two Hubbard sub-bands, which is an important signature of correlated hopping. This asymmetry is distinct from that due to an asymmetric coupling between the dot and the external electrodes. The latter asymmetry was found to be crucial for the rectification of both longitudinal and perpendicular currents in the four terminal cross geometry~\cite{kiw2023}.
    
The paper is organized as follows. In Section \ref{sec:model} we present the model with correlated-hopping term and describe the approach. The transport characteristics of interest in the strongly non-linear regime are introduced in Section \ref{sec:nltr-coeff}. The results for the conductance and the Seebeck coefficient are presented and discussed in Section \ref{sec:res}, and we end up with summary and conclusions in Section \ref{sec:sum-concl}.

\section{The model and basic definitions}
\label{sec:model}
The present study expands our previous work into non-linear regimes. The goal is to identify the signatures of the correlated-hopping interaction in the transport characteristics. This section recapitulates the main points, discussed earlier in detail~\cite{eckern2021}.

\subsection{General aspects}
\begin{figure}
\includegraphics[width=1.0\linewidth]{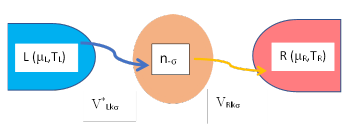}
\caption{(Color online) An electron with spin $\sigma$ and wave vector $k$ jumps from one of the electrodes (here left) onto the quantum dot with amplitude $V^\star_{Lk\sigma}$ (or $V_{Lk\sigma}$ for opposite jump). In the model with correlated hopping the amplitude of the process depends on the existence of opposite spin electron on the quantum dot. In the Hamiltonian (\ref{eq:ham1}) this is conveniently represented by the composite annihilation operator $D_\sigma=d_\sigma (1-xn_{\bar{\sigma}})$ and its Hermitian conjugate. }
\label{fig:rys100}
\end{figure}

The Hamiltonian of the studied system, schematically shown in Fig.~\ref{fig:rys100}, with correlated hopping, is similar to the standard single-impurity Anderson model, albeit with modified on-dot operators that take the correlated hopping into account
\beq
{H}&=&\sum_{\lambda {k} \sigma}\varepsilon_{\lambda {k}}n_{\lambda {k} \sigma} + 
\sum _{\sigma} \varepsilon _{\sigma}n_{\sigma} +Un_\uparrow n_\downarrow \nonumber \\
&+&\sum_{\lambda {k} \sigma} \left({V}_{\lambda {k}\sigma} c^{\dagger}_{\lambda k\sigma} D_{\sigma} 
+ {V}_{\lambda {k}\sigma}^* D^{\dagger}_{\sigma} c_{\lambda {k} \sigma}\right),
\label{eq:ham1}
\eeq
where the composite operator $D_\sigma=d_\sigma(1-xn_{\bar{\sigma}})$ takes care of the occupation dependence of the hopping. Number operators $n_{\lambda {k} \sigma}=c^{\dagger}_{\lambda {k} \sigma}c_{\lambda {k} \sigma}$~and $n_{\sigma}=d^{\dagger}_{\sigma} d_{\sigma}$ are defined for the leads and the dot, respectively.  The operators $c^{\dagger}_{\lambda k\sigma} (d^{\dagger}_{\sigma})$ create electrons in respective states $\lambda {k}\sigma$ $(\sigma)$ in the lead $\lambda$ (on the dot). The energies of the leads are measured from their chemical potentials $\mu_\lambda$, so $\varepsilon_{\lambda k}=\varepsilon_{0\lambda k}-\mu_\lambda$, with the dependence of $\varepsilon_{0\lambda k}$ on $\lambda$ allowing for a different spectrum in each of the leads. The spin quantum number $\sigma=\pm 1$ are for $\uparrow, \downarrow$ spins, respectively. 
The standard hopping term from the quantum dot (QD) to the electrode $\lambda$ is written as $\sum_{\lambda {k} \sigma} ({V}_{\lambda {k}\sigma} c^{\dagger}_{\lambda k\sigma} d_{\sigma}+H.c.)$, while the state-dependent hopping has been parameterized by the parameter $x$ which is minus the ratio between $K_{\lambda k \sigma}$ and ${V}_{\lambda k\sigma}$, $x= - {K_{\lambda k \sigma}}/{V_{\lambda k \sigma}}$. We assume $x$ to be constant~\cite{eckern2021}, i.e., not dependent on $\lambda k \sigma$. The on-dot energy depends on the applied magnetic field $B$ as $\varepsilon_\sigma=\varepsilon_d+\sigma \mu_B B$, where $\mu_B$ is the Bohr magneton and $\varepsilon_d$ the bare on-dot electron energy level. 
The latter parameter can be tuned by an external gate voltage (not shown in Fig.~\ref{fig:rys100}) which couples to the quantum dot. We shall also use $\delta=\varepsilon_d+U/2$, which is the departure from the particle-hole symmetry point. The Hubbard parameter $U$ describes the repulsion between two electrons on the dot. For the particle-hole symmetric system $\delta=0$ if $x=0$. A non-zero value of $x$ breaks time-reversal symmetry.

For completeness, we note that in addition to $U=\langle ii|V|ii\rangle$ and $K_{ij}=\langle ii|V|ij\rangle$, where $i$ refers to a dot state and $j$ to a lead state, there are other Coulomb terms, related to the amplitudes $\langle ij|V|ij\rangle$, $\langle ii|V|jj\rangle$, and $\langle ij|V|ji\rangle$; c.f.\ the quite explicit discussion in Ref.~\onlinecite{hubbard1963}. Quantitative estimates, of course, depend on the actual nature of the states involved, but considering the exponential factors arising due to the localized nature of the Wannier states, the following inequalities can be expected~\cite{hubbard1963}:
$U \gg K \gg \langle ii|V|jj\rangle \approx \langle ij|V|ji\rangle$. On the other hand, the contributions arising from $\langle ij|V|ij\rangle$ can be quite appreciable in magnitude, leading to terms $\propto n_i n_j$ (recall that $i$ is on the dot and $j$ in a lead), but still small compared to $U$; these are proportional to the dot density, and thus are not expected to have a pronounced effect on the transport across the dot. Of course, the discussion of Coulomb interaction terms depends on the concrete system under study, and has to be modified when the dot and/or the leads are more complex, or when the spin degree of freedom is relevant; see, e.g., Ref.~\onlinecite{ciorga2000}.

Below we use the standard approach~\cite{wingreen1994} and calculate the charge current
flowing out of the electrode $\lambda$ as ($e$ times) the time derivative of the average particle number, $\langle N_{\lambda}\rangle=\sum_{k\sigma} \langle n_{\lambda{k}\sigma}\rangle$, of lead $\lambda$. 
Application to the two-terminal QD we are interested in here provides  $I=I_L=-I_R$, which expresses current conservation in the system: 
\be
I=\frac{2e}{\hbar} \sum_\sigma \tilde{\Gamma}_\sigma\int\frac{dE}{2\pi}\left[f_L(E)-f_R(E)\right]\mathrm{Im} G^r_\sigma(E), 
\label{eq:c-curr-wbl}
\ee
with $\tilde{\Gamma}_\sigma={\Gamma_\sigma^L\Gamma_\sigma^R}/\left({\Gamma_\sigma^L+\Gamma_\sigma^R}\right)$; the parameters
\be
\Gamma_\sigma^{\lambda}(E)=2\pi\sum_{{k}}|V_{\lambda{k}\sigma}|^2\delta(E-\varepsilon_{\lambda{k}})
\label{eq:gammas}
\ee
describe the coupling between the dot and the electrode. They are assumed to be independent of energy $E$, which corresponds to the wide-band limit. It is important to stress that the wide-band limit condition is not always valid: notable examples are hybrid systems \cite{polkovnikov2002,wysokinskimm2013,wysokinskimm2014} with one (or both) of the electrodes being a superconductor. 
 
We emphasize that the model described by the Hamiltonian (\ref{eq:ham1}) requires the knowledge of two different Green functions (GFs): the transport GF defined as   
\be
G^r_\sigma(E)=\langle\langle  D_\sigma|D^\dagger_\sigma\rangle\rangle^r_{E},
\label{eq:gf-transport}
\ee
which enters the formulae for the currents, and the spectral GF
\be
g^r_\sigma(E)=\langle\langle  d_\sigma|d^\dagger_\sigma\rangle\rangle^r_{E},
\label{eq:gf-spectral}
\ee
 describing {\it inter alia} the occupation $\langle n_\sigma\rangle$ of the quantum dot. 

The Fermi function 
$$
f_{L/R}(E)= \left[ \exp\frac{(E-\mu_{L/R})}{k_BT_{L/R}}+1 \right]^{-1}
$$
describes the electrons' distributions in the leads $L/R$, assumed to be in local equilibrium at temperature $T_{L/R}$ and chemical potential $\mu_{L/R}$, respectively. 

\subsection{Comments on GFs, self-energies and lifetimes}
The GFs (\ref{eq:gf-transport}) and (\ref{eq:gf-spectral})
have been calculated previously~\cite{eckern2021}, and we shall not repeat the details here. The final formulae can be found in the Supplemental Material~\cite{suppl}, together with the derivation of the fourth-order lifetime effects. Instead, we recall a few important aspects only. The general structure of both GFs is the following:
\be
F_\alpha(E)=\frac{1+n_\alpha(E)}{E-\varepsilon_d-\Sigma_{0\sigma}(E)-\Sigma_\alpha(E)}.
\ee
Here we briefly use the symbol $F_\alpha(E)$, with $\alpha = \mathit{transport}$ or $\alpha = \mathit{spectral}$ to denote the two GFs. The self-energy
\beq
\Sigma_{0\sigma} (E)=\sum_{\lambda k} \frac{|V_{\lambda k \sigma}|^{2}}{E-\varepsilon_{\lambda k}}
\label{sigma0}
\eeq
results from the coupling to the leads. In the wide-band limit one approximates (\ref{sigma0}) by its imaginary part:
\beq
\Sigma_{0\sigma} (E)&\approx& -i\pi \sum_{\lambda k}{|V_{\lambda k \sigma}|^{2}}\delta(E-\varepsilon_{\lambda k}) \nonumber \\
&=&-i(\Gamma^L_\sigma+\Gamma^R_\sigma)/2 
\label{Gam-wbl}
\eeq

The dimensionless functions $n_\alpha(E)$ depend on the average occupation of the dot $\langle n_{\bar{\sigma}}\rangle$, the parameter $x$, interaction $U$, and other parameters of the system. These functions, as also the self-energies $\Sigma_\alpha(E)$, are sums of various self-energy pieces $b_{i{\bar{\sigma}}}$,
${\bar b}_{i{\bar{\sigma}}}$, and ${\tilde b}_{i{\bar{\sigma}}}$,
which in turn depend on the GFs in a self-consistent fashion. 
As examples, we quote the formulae
\beq
\tilde{b}_{1\bar{\sigma}} (E)=\sum_{\lambda k}\frac{V^{*}_{\lambda k \bar{\sigma}}
\langle D^{\dagger}_{\bar{\sigma}} c_{\lambda k \bar{\sigma}}\rangle}{E-\varepsilon_{\lambda k}-\varepsilon_1 +i\tilde{\gamma}^{\bar{\sigma}}_1} \nonumber \\
=\int \frac{d\varepsilon}{2\pi} \frac{\sum_{\lambda}\Gamma^{\lambda}_{\bar{\sigma}} f_{\lambda} (\varepsilon)\langle\langle D_{\bar{\sigma}} |D^{\dagger}_{\bar{\sigma}}\rangle\rangle ^{a}_{\varepsilon}}{E-\varepsilon 
-\varepsilon_1 +i\tilde{\gamma}^{\bar{\sigma}}_1},
\eeq
where $\varepsilon_1=\varepsilon_\sigma-\varepsilon_{\bar{\sigma}}$ denotes the energy of a singly occupied spin polarized  state, and
\beq
b_{2\bar{\sigma}}(E)
= \sum_{\lambda k}\frac{V_{\lambda k\bar{\sigma}}\langle c^{\dagger}_{\lambda k \bar{\sigma}}d_{\bar{\sigma}}\rangle}
{E+\varepsilon_{\lambda k}-\varepsilon_2+ i \tilde{\gamma}_2} \nonumber \\
= \int\frac{d\varepsilon}{2\pi} \frac{\sum_{\lambda} \Gamma^{\lambda}_{\bar{\sigma}} 
f_{\lambda}(\varepsilon) \langle\langle d_{\bar{\sigma}} |D^{\dagger}_{\bar{\sigma}}\rangle\rangle^{r}_{\varepsilon}}
{E +\varepsilon -\varepsilon_2+i\tilde{\gamma}_2},
\eeq
where $\varepsilon_2=\varepsilon_\sigma+\varepsilon_{\bar{\sigma}}+U$ is the energy of the doubly occupied quantum dot.
Note that the above expressions are calculated with inverse lifetimes $\tilde{\gamma}_1^{\bar{\sigma}}$ and $\tilde{\gamma}_2$ which replace the $0^+$ symbols of the standard approach, and approximately take care of higher order processes~\cite{lavagna2015}. The explicit expressions for them are given in Ref.~\onlinecite{suppl}. Needless to say, the lifetimes affect the numerical values of all transport coefficients. This is obvious as they are a source of the so-called lifetime broadening of quantum states. 

\begin{figure}
\includegraphics[width=0.95\linewidth]{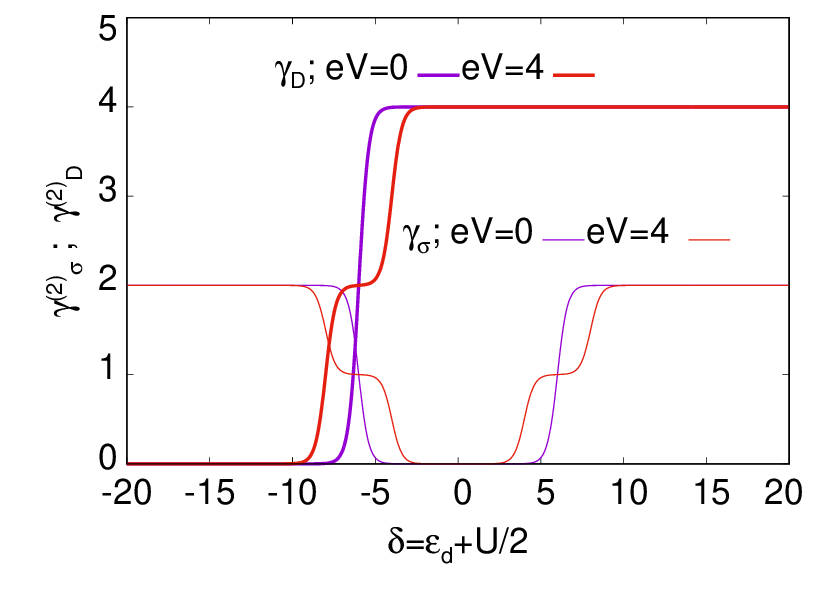}
\caption{(Color online) Dependence of second order inverse lifetimes on $\varepsilon_d$ for two values of voltage bias, $eV=0$ and $eV=4$. The other parameters are $U=12$, $T=0.3$. At lower temperatures, the changes would be very steep. Note that a doubly occupied state is short-lived for all $\varepsilon_d > -U$. Here we assume that all $\Gamma$'s are equal to each other (i.e., the $L$-$R$-couplings are the same and independent of spin), which we take as energy unit; also we consider, here and below, the case $B=0$.
}
\label{fig:rys20a}
\end{figure}
The inverse lifetimes are calculated by the Fermi rule up to second order in the couplings (\ref{eq:gammas}), or fourth order in the amplitudes $V_{\lambda k \sigma}$. For a singly occupied state, the inverse lifetime in principle is spin dependent, and denoted by
$\tilde{\gamma}^{\sigma}_1=\gamma^{(2)}_\sigma+\gamma^{(4)}_\sigma$, while
$\tilde{\gamma}_2=\gamma^{(2)}_D+\gamma^{(4)}_D$. We obtain
\beq
\gamma^{(2)}_\sigma&=&\sum_{\lambda}\left(\Gamma^\lambda_\sigma [1-f_\lambda (\varepsilon_\sigma)]
+(1-x)^2\Gamma^\lambda_{\bar{\sigma}} f_\lambda (\varepsilon_{\bar{\sigma}}+U)\right),
\nonumber \\
\gamma_D^{(2)}&=&(1-x)^2\sum_{\lambda \sigma}\Gamma^\lambda_\sigma [1-f_\lambda (\varepsilon_\sigma+U)],
\eeq
and 
\beq
\gamma^{(4)}_\sigma&=&\frac{1}{2\pi}\sum_{\lambda\lambda'}\int d\varepsilon[1-f_\lambda(\varepsilon)]\left[\Gamma^\lambda_\sigma \Gamma^{\lambda'}_{\bar{\sigma}}f_{\lambda'}(\varepsilon+\varepsilon_{\bar{\sigma}}-\varepsilon_\sigma) \right.  \nonumber \\
&\times& \left. \left(\frac{(1-x)^2}{\varepsilon-\varepsilon_\sigma-U}-\frac{1}{\varepsilon-\varepsilon_\sigma}\right)^2 \right.  \nonumber \\
&+& \left.f_{\lambda'}(\varepsilon)\left(\frac{\Gamma^\lambda_\sigma\Gamma^{\lambda'}_\sigma}{(\varepsilon-\varepsilon_\sigma)^2}+\frac{(1-x)^4\Gamma^\lambda_{\bar{\sigma}}\Gamma^{\lambda'}_{\bar{\sigma}}}{(\varepsilon-\varepsilon_{\bar{\sigma}}-U)^2}\right)\right],
\label{eq:4sigma}
\eeq
 and 
\beq
\gamma^{(4)}_D&=&\frac{(1-x)^2}{2\pi}\sum_{\lambda\lambda'\sigma}\int d\varepsilon\frac{1-f_\lambda(\varepsilon)}{(\varepsilon-\varepsilon_\sigma-U)^2}\left[(1-x)^2 \Gamma^\lambda_\sigma\Gamma^{\lambda'}_\sigma\times  \nonumber \right.\\
&\times&f_{\lambda'}(\varepsilon)+\left.\Gamma^\lambda_\sigma\Gamma^{\lambda'}_{\bar{\sigma}}[1-f_{\lambda'}(\varepsilon_\sigma+\varepsilon_{\bar{\sigma}}+U-\varepsilon)]\right].
\label{eq:4D}
\eeq
For details see the Supplemental Material~\cite{suppl}. Before proceeding, let us note that the inverse lifetimes are non-monotonous functions of the on-dot energy $\varepsilon_d$. This is well visible in Fig.~\ref{fig:rys20a}, showing the lowest order contributions $\gamma^{(2)}_\sigma$ and $\gamma^{(2)}_D$. The (abrupt at low temperatures) changes of inverse lifetimes may lead to small jumps of the conductance as function of gate voltage $\delta$. The inverse lifetime $\gamma^{(2)}_D$ outside the energy window $|\varepsilon|<U/2$ additionally does not preserve particle-hole symmetry even for $x=0$, as is apparent in Fig.~\ref{fig:rys20a}; the same is true for $\gamma^{(4)}_D$ (not shown). This leads to a small, albeit visible asymmetry between the conductance peaks corresponding to singly and doubly occupied dot states.
These two lifetime-related small features are discernible in the figures below but will not be discussed each time they appear.

Before proceeding we comment on the numerical procedure. The set of equations for both Green functions~\cite{suppl} is solved by iteration, and the iteration proceeds until convergence is reached. In each step, we check the actual occupation of the dot $\langle n_\sigma\rangle$, and the iteration is terminated when the difference is less than 0.0001. To deal with the divergent terms in the inverse lifetimes, Eqs.~(\ref{eq:4sigma}) and (\ref{eq:4D}), we cut the divergence each time a denominator is closer to zero than 0.16.

\section{Conductance and Seebeck coefficient in the non-linear regime} 
\label{sec:nltr-coeff}
As the charge current depends both on the voltage and temperature bias, one may define several kinetic and transport coefficients. For small external forces $V \to 0$ and $\Delta T \to 0$, the linear expansion is valid so we have $I(\Delta T,V)=L_{11}V+L_{12}\Delta T$. This shows that the conductance defined as the linear response of the thermally homogeneous system ($\Delta T=0$) to the applied voltage $I(V)=GV$, is given as $G=L_{11}$. On the other hand, the thermopower, characterized by the Seebeck coefficient
\begin{equation}
S=-\left(\frac{V}{\Delta T}\right)_{I=0, \Delta T \to 0},
\label{seebeck}
\end{equation}
is defined under the condition of zero charge current.
In the linear regime, $S$ is given by the ratio of kinetic coefficients,
$S=L_{12} / L_{11}$. Expanding Eq.~(\ref{eq:c-curr-wbl}) for small $V$ and $\Delta T$ up to linear order, one finds an explicit expression for $S$.

The above formal definitions of conductance and Seebeck coefficient can be used for bulk systems and 
nano-structures with two external electrodes~\cite{costi2010}. However, they have to be generalized for multi-terminal devices as discussed by B\"uttiker~\cite{buttiker1986} in the context of the conductance and generalized by Butcher~\cite{butcher} and others~\cite{mazza2014,michalek2016} in the context of the thermopower, where matrices of conductances, respectively Seebeck coefficients, have been introduced.

Strong non-linear effects are expected to occur in nano-devices \cite{benenti2017,zimbovskaya2011}. Hence in such systems, appropriate definitions of the conductance and thermoelectric coefficients are required. The  formula (\ref{eq:c-curr-wbl}) for the current is valid for arbitrary voltages $V=(\mu_R-\mu_L)/e$, where $e$ is the elementary charge, and arbitrary temperature difference $\Delta T$. As the generalization of the standard Ohm's law, we define the differential conductance
\be
G_d (V)=\left(\frac{\partial I(V)}{\partial V}\right)_{\Delta T=0}.
\label{cond-dif}
\ee 
Similarly, the definition of the Seebeck coefficient can be extended to the differential one, $S_d$, calculated formally for constant current flowing as a result of the external voltage $V$. This $S_d$ measures the response of the system with the current flow to a minute change in temperature difference. At a given applied external voltage $V$ and temperature difference $\Delta T$, $S_d$ is hence defined \cite{dorda2016,manaparambil2023,daroca2018} as 
\begin{equation}
S_d=-\left(\frac{\partial V}{\partial \Delta T}\right)_{I} 
=-\left(\frac{\partial I}{\partial \Delta T}\right)_V\bigg/\left(\frac{\partial I}{\partial V}\right)_{\Delta T}.
\label{seebeck-dif}
\end{equation}
As argued earlier, $S_d$ should be accessible experimentally \cite{dorda2016} in appropriate AC circuits.
The definition (\ref{seebeck-dif}) is analogous to the differential conductance 
(\ref{cond-dif}) usually employed in the non-linear regime.

Another suitable generalization of the definition (\ref{seebeck}) of the Seebeck coefficient of the two-terminal systems to non-linear situations has been considered~\cite{eckern2020}: 
\begin{equation}
S_n=-\left(\frac{V}{\Delta T}\right)_{I(\Delta T, V)=0}.
\label{seebeck-2}
\end{equation} 
This definition is in accord with the traditional way of measuring the Seebeck coefficient. One applies a temperature bias $\Delta T$ and determines the voltage $V$ such that the current across the device vanishes, $I(\Delta T, V)=0$. In this formulation, the only source of non-linearity is directly given by the value of $\Delta T$, which is large and precludes the linear expansion of the charge current $I(\Delta T,V)$ in first powers of $V$ and $\Delta T$. 

Of course, for infinitesimally small $\Delta T$ and $V$ all definitions are equivalent and lead to the same result, $S=S_n=S_d$. For arbitrary $V$ and $T$ but vanishingly small $\Delta T$, the two non-linear Seebeck coefficients are equal, $S_n=S_d$. For large $\Delta T$ strong non-linearities are expected, and the different definitions lead to quantitative differences between the Seebeck coefficients as discussed in detail earlier~\cite{eckern2020}. 
In the next section, we shall present results for the differential conductance, $G_d$, and the differential Seebeck coefficient, $S_d$.

\section{Results}
\label{sec:res}
We concentrate on the conductance and the Seebeck coefficient as these are easiest to access experimentally. Their measurement as functions of the gate voltage (or, equivalently, of the position of the on-dot energy level $\varepsilon_d$, or $\delta$) provides interesting information. In an experiment, one continuously tunes the occupation of the system which, with changing gate voltage, changes from empty to singly and doubly occupied states or the other way around. In studying quantum dots, as modelled by the Hamiltonian (\ref{eq:ham1}), most parameters can be controlled except $x$. In all figures presented below, we consider $B=0$ and assume that the $L/R$ coupling parameters are spin independent, hence dropping the spin subscript in the following, $\Gamma^{L/R}_\sigma \to \Gamma^{L/R}$. In addition, we take $\Gamma^L$ as the energy unit.

\subsection{Symmetric coupling: $\Gamma^L = \Gamma^R$}
\label{subsec:res-symmetric}
As discussed earlier, the numerical values of transport coefficients do depend on the lifetimes. For a model with $x=0$, the differential conductance $G_d(\delta)$ is a symmetric function of $\delta$ for all values of voltage $V$.  On the contrary, but in agreement with expectations, the Seebeck coefficient $S_d$ is an antisymmetric function of $\delta$. This property does not depend on the voltage bias or its distribution if $x=0$ and the coupling constants are symmetric, $\Gamma^L=\Gamma^R$. In the following, we always assume a symmetric distribution of voltages and temperatures with $\mu_{R/L}=\mu \pm eV/2$ and $T_{L/R}=T\pm \Delta T/2$, with $T$ the common temperature of the system.

\begin{figure}
\includegraphics[width=0.95\linewidth]{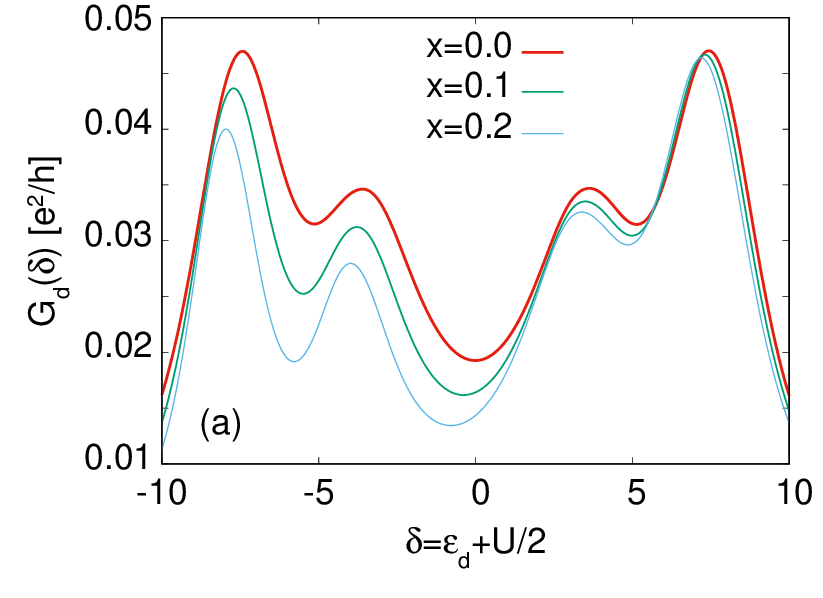}
\includegraphics[width=0.95\linewidth]{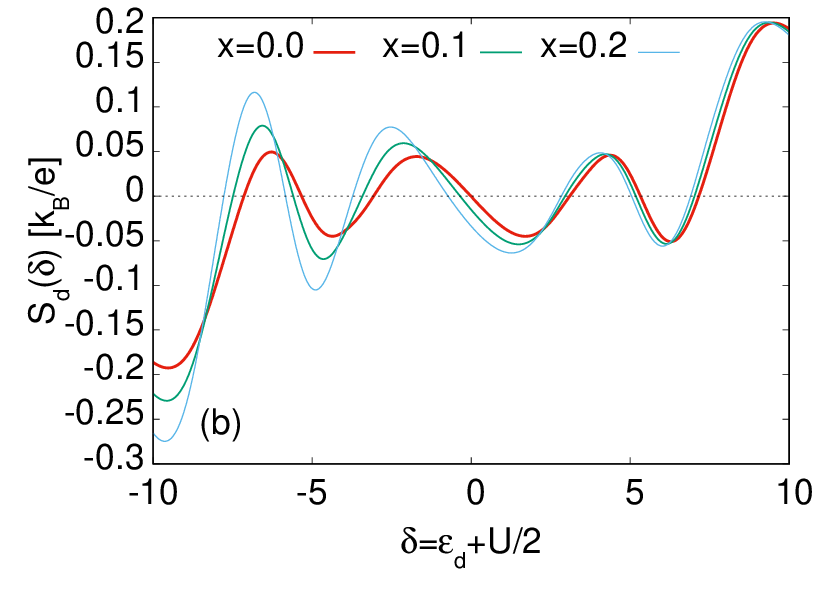}
\caption{(Color online) The dependence of the differential conductance (a) and the Seebeck coefficient (b) on the gate voltage, parameterized by $\delta$, for $eV=4$ and two values of $x$, 0.1 and 0.2. The other parameters are $U=12$ and $T=0.3$. The energies and the temperature are expressed in units of $\Gamma^L (= \Gamma^R)$.
}
\label{fig:rys1b}
\end{figure}

To elucidate the influence of correlated hopping on the conductance and the thermopower in a strongly non-equilibrium situation, in Fig.~\ref{fig:rys1b} we present both transport characteristics calculated for the voltage $eV=4$ and two values of $x$, 0.1 and 0.2. For comparison, we also plot the curve for $x=0$. Importantly, with $x\ne 0$, we observe a breaking of particle-hole symmetry, which results in the well-visible shift of the minimum of conductance towards negative $\delta$, accompanied by a lowering of the overall conductance peak and its splitting at negative $\delta$. At gate voltages corresponding to positive $\delta$, on the contrary, one observes only minute changes in the conductance peak line shape. Even though the lack of particle-hole symmetry is well visible also in the linear transport as discussed earlier~\cite{eckern2021}, here we observe additional asymmetries. Namely, the splitting of the lower conductance band strongly increases with $x$ in comparison with the splitting of the upper one, which remains the same as for $x=0$. This is an important additional characteristic detail that allows the identification of the effect of correlated hopping, $x\ne 0$, in the non-linear conductance spectra.

The differential thermopower $S_d(\delta)$ also features interesting correlated-hopping induced differences between low and high values of $\delta$. In particular, these differences are much larger at negative $\delta$ values. This goes in line with the decrease in the width of the split conductance peak.   
 \begin{figure}
\includegraphics[width=0.95\linewidth]{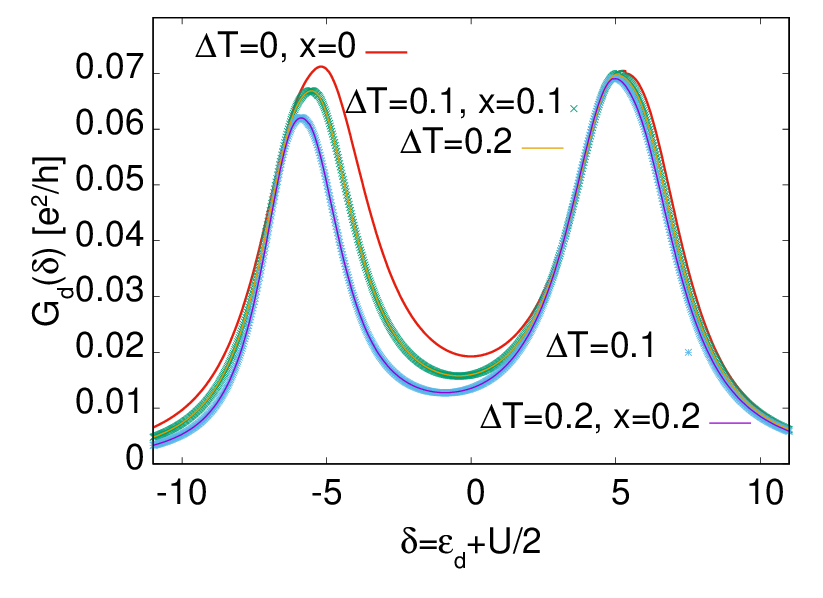}
\caption{(color online) The dependence of the differential conductance $G_d$ on $\delta$ for $eV=0$, for two values of the thermal bias $\Delta T$ (0.1, 0.2) symmetrically distributed $T_{L/R}=T\pm\Delta T/2$, and for two values of $x$, 0.1 and 0.2. For comparison, the curve with $x=0$ and $\Delta T=0$ is also plotted. The other parameters are $U=12$ and $T=0.3$. Note that on the scale of the figure, the effect of thermal bias is not visible, i.e., the curves for the same $x$ and different $\Delta T$ overlap.}
\label{fig:rys1c}
\end{figure} 
\begin{figure}
\includegraphics[width=0.95\linewidth]{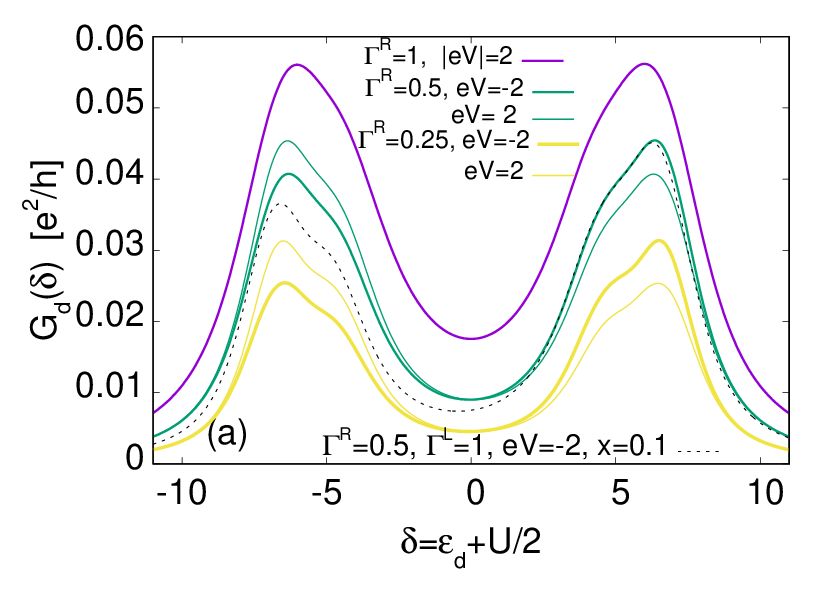}
\includegraphics[width=0.95\linewidth]{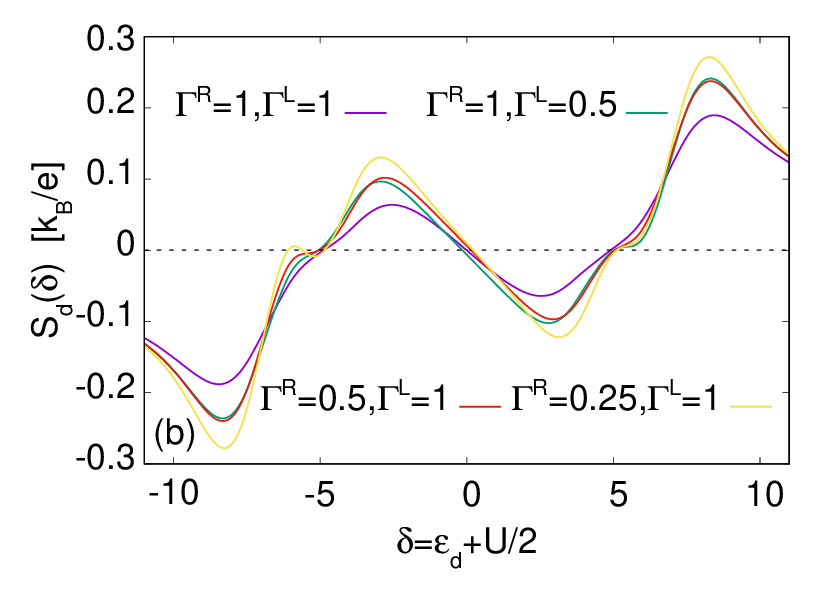} 
\caption{(Color online) The effect of asymmetric couplings, $\Gamma^R\ne\Gamma^L$, on the differential conductance (a), and on the differential thermopower (b), as functions of $\varepsilon_d$. In panel (a), the curves of the same colour differ by the direction of voltage $eV=+2$ vs. $eV=-2$, 
hence they characterize the rectification properties of the system with broken inversion symmetry. The purple curve is for the symmetric system with $\Gamma^R=\Gamma^L$ for which the conductance does not depend on the sign of the bias, i.e., is the same for $eV=\pm 2$. The other parameters are $U=12$, $T=0.3$, and $x=0$ for all curves, except for the dotted line in panel (a) which is obtained with $x=0.1$.}
\label{fig:rys3}
\end{figure}

One also observes the changes of the conductance calculated for thermal bias $\Delta T$. The effect is shown in Fig.~\ref{fig:rys1c} where we present $G_d(\delta)$ for $x=0, 0.1, 0.2$, and two values of the temperature difference $\Delta T=0.1, 0.2$, with the overall temperature $T=0.3$. The effect of $\Delta T$ on the conductance for a given value of $x$ is very small; in fact, invisible on the scale of Fig.~\ref{fig:rys1c}. However, for $x\ne 0$, we observe a clear breaking of particle-hole symmetry and the shift of the minimum conductivity towards negative values of $\delta$, while the thermopower does not visibly change with $\Delta T$, i.e., the curves for the same $x$ but different $\Delta T$ overlap.

\subsection{Asymmetric coupling: $\Gamma^L \neq \Gamma^R$}
\label{subsec:res-asymmetric}
The asymmetry of the couplings $\Gamma^L$, $\Gamma^R$ between the dot and the electrodes also leads to modifications of the line shapes of conductance and thermopower, and can thus mask the effect of $x$. It turns out, however, that these changes are markedly distinct. Disparate couplings between the dot and the two electrodes break the inversion symmetry of the system and result in rectification effects~\cite{kiw2023}, which typically show up as the dependence of the current on the voltage bias direction: $I(V)\ne I(-V)$. Here, the rectification effects are seen as a difference between $G_d(\delta, V)$ and $G_d(\delta,-V)$. 

We illustrate rectification and also the effect of the coupling asymmetry on the conductance in Fig.~\ref{fig:rys3}a, where we change $\Gamma^R$ leaving $\Gamma^L$ ($= 1$) fixed. In the figure, we also show the conductance for the symmetric system, $\Gamma^R=\Gamma^L$, when there is no rectification and the conductance is the same for $eV=\pm 2$. On the contrary, for $\Gamma^R=0.5$ or 0.25 shown by green or yellow lines, respectively, the conductances for positive voltage (thin lines) are markedly different from those for negative voltage (thick lines of the same colour). The dashed line corresponds to $\Gamma^R=0.5$ and $eV=-2$ (with the same couplings as for a green thick line) but for $x=0.1$. The effects of the asymmetry of the couplings are seen to be markedly different from those related to $x$. As is apparent in the figure, the conductance for $x=0$ fulfills the general symmetry $G_d(\delta, V)=G_d(-\delta,-V)$ which can be easily checked experimentally. Departures from this symmetry would uniquely indicate the presence of correlated hopping, i.e., non-zero $x$.

The difference between $G_d(\delta,x, V)$ and $G_d(\delta,x,-V)$, discernable in Fig.~\ref{fig:rys3}a, is due to the rectification effect existing in systems with broken inversion symmetry. In such a situation there is no obvious relation between conductance line shapes measured for positive vs.\ negative $\delta$, or forward vs.\ backward voltage bias. An unequivocal signature of correlated hopping is the shift of the minimum between the two conductance bands towards negative values of $\delta$. Finally, we mention that there exists still another symmetry related to the $x$ parameter: for each $\delta$ and arbitrary $V$ one finds $G_d(\delta, x)=G_d(\delta, 2-x)$, resulting from the general symmetry of the Hamiltonian~\cite{eckern2021}. 

In Fig.~\ref{fig:rys3}b, we show $S_d(\delta)$ for fixed voltage $eV=+2$, and for varying couplings. The purple line for $\Gamma^L=\Gamma^R$ shows the antisymmetric Seebeck coefficient ($S_d(\delta,V)=-S_d(-\delta,V)$. Green and red lines illustrate the rectified thermopower obtained for $\Gamma^R=1$, $\Gamma^L=0.5$ (green line) and $\Gamma^R=0.5$, $\Gamma^L=1$ (red curve). Such an alteration of couplings is equivalent to reversing the voltage direction in the device with unchanged couplings. The yellow line corresponds to an even higher anisotropy, with $\Gamma^L=1$, and $\Gamma^R=0.25$. 

\subsection{Low temperatures}
\label{subsec:res-lowtemp}
\begin{figure}
\includegraphics[width=0.95\linewidth]{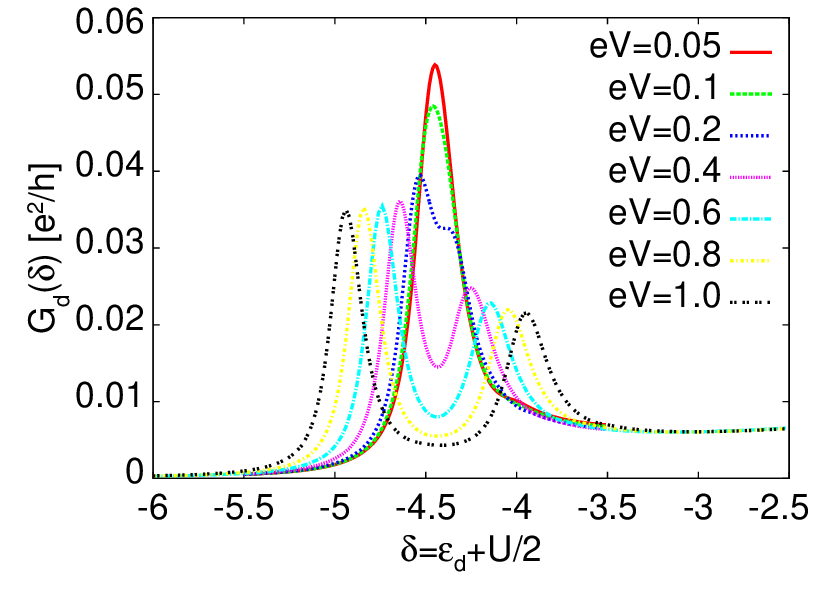}
\caption{(Color online) The dependence of the differential conductance on the gate voltage, parameterized by $\delta$, for several $eV$ values, as indicated, and for symmetric coupling, $\Gamma^L=\Gamma^R$. Here $x=0.8$, and the other parameters read $U=8$, $T=0.02$.}
\label{fig:rys0}
\end{figure}

The method employed here for calculating Green functions is also able to qualitatively describe the formation of the Kondo resonance at low temperatures, as discussed earlier~\cite{lavagna2015,eckern2021}. Hence, we next investigate this regime and show the differential conductances and Seebeck coefficients for very low temperatures, and a few values of the interaction strength $U$. In Fig.~\ref{fig:rys0}, we present the splitting of the lower conductance peak by the bias voltage $V$, for a system with $U=8$, $x=0.8$, $T=0.02$, and symmetric coupling $\Gamma^R=\Gamma^L$; the inverse lifetimes are approximated by their lowest order values, i.e., neglecting fourth order contributions:
$\tilde\gamma^\sigma_1 \simeq \gamma^{(2)}_\sigma$, $\tilde\gamma_2 \simeq \gamma^{(2)}_D$.
One observes a clear splitting of the lower conductance peak for $eV \gtrsim 0.3$.

\begin{figure}
\includegraphics[width=0.95\linewidth]{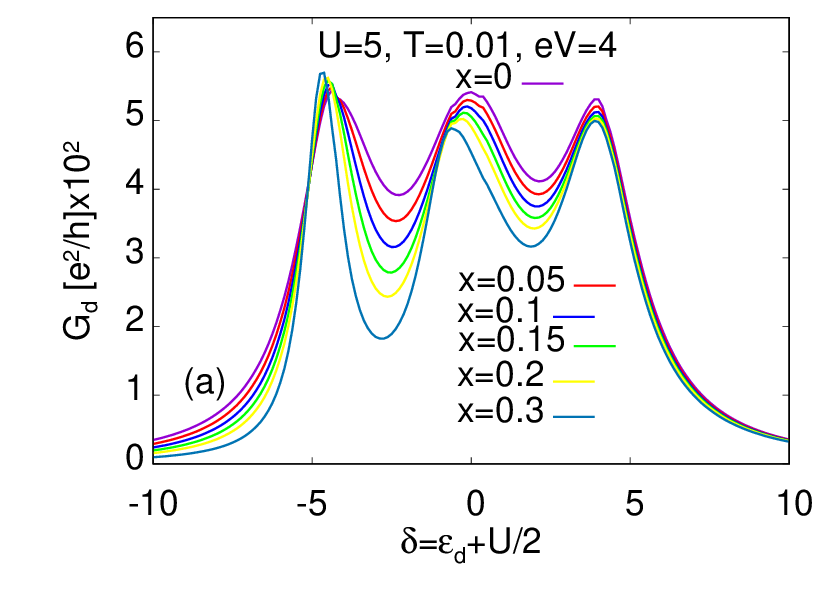}
\includegraphics[width=0.95\linewidth]{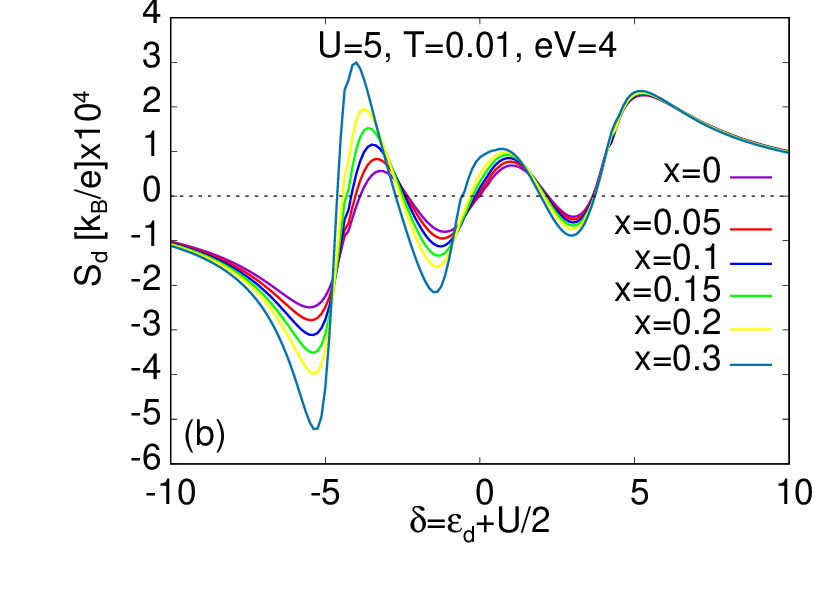}
\caption{(Color online) The conductance (upper panel) and the Seebeck coefficient (lower panel) vs.\ gate voltage, parameterized by $\varepsilon_d$, for several $x$ values, as indicated, and for symmetric coupling, $\Gamma^L=\Gamma^R$. The other parameters read $eV=4$, $U=5$, and $T=0.01$.}
\label{fig:rys1}
\end{figure}

\begin{figure}
\includegraphics[width=0.95\linewidth]{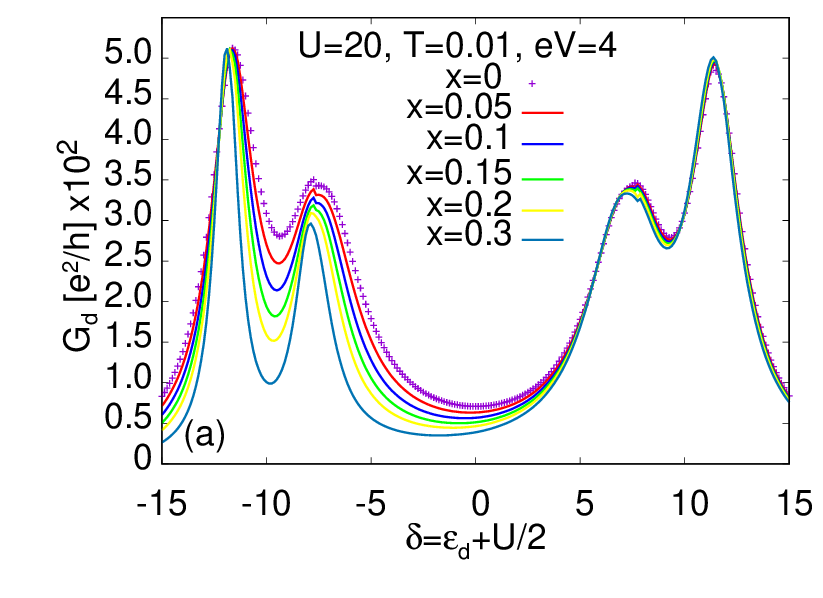}
\includegraphics[width=0.95\linewidth]{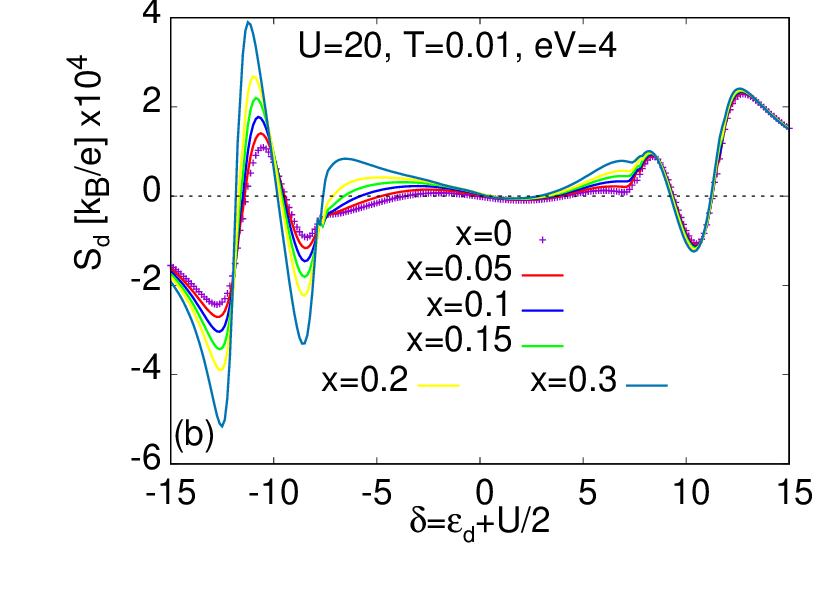}
\caption{(Color online) The dependence of the conductance (upper panel) and the Seebeck coefficient (lower panel) on the gate voltage parameterized by $\varepsilon_d$ for a number of $x$ values, as indicated, and for symmetric coupling, $\Gamma^L=\Gamma^R$. The other parameters are the same as in the previous figure except $U=20$.}
\label{fig:rys2}
\end{figure}

Next, Figs.~\ref{fig:rys1} and \ref{fig:rys2}, we discuss the conductances (upper panels) and Seebeck coefficients (lower panels) as functions of the gate voltage, i.e., vs.\ $\delta$, for a fixed relatively large voltage, $V$ ($eV=4$). In Fig.~\ref{fig:rys1}a it is seen that for relatively moderate $U$ ($U=5$) the conductance features only three instead of four maxima. At the particle-hole symmetric point $\delta=0$ (for $x=0$; recall that $x=0$ corresponds to the standard Hubbard model) one obtains a maximum in the conductance and the concomitant vanishing of the thermopower. The small features around the conductance maxima are related to the Kondo resonance, and to the modifications of the lifetimes mentioned earlier (c.f.~Fig.~\ref{fig:rys20a}). The two maxima in the conductance corresponding to the lower and upper Hubbard bands are located at $\delta \simeq \pm 4$. With increasing $x$ the left maximum moves towards negative $\delta$ and the peak gets narrower. As a consequence, the left minimum is strongly modified. One observes its deepening accompanied by an increase of its width. The $x$-induced changes in the conductance are less pronounced in the middle maximum, and almost absent in the right maximum.

Analogous changes are observed in the Seebeck coefficient shown in Fig.~\ref{fig:rys1}b, which features three minima and three maxima. An increase of $x$ hardly affects the rightmost maximum, while the left part of $S_d(\delta)$ is strongly modified. The peaks of the thermopower at negative $\delta$ are higher and narrower in comparison to those for positive $\delta$. We underline that for $x=0$ both the conductance and the Seebeck coefficient are symmetric, respectively anti-symmetric, functions of $\delta$. These modifications of the line shapes are unique signatures of correlated hopping. Note that the considered temperature, $T=0.01$, is much lower than the Kondo temperature $T_K$ of the particle-hole symmetric model ($x=0$, $\varepsilon_d=-U/2$). 

A similar behavior is observed in Fig.~\ref{fig:rys2}, with even more pronounced changes of the line shapes of $G_d(\delta)$ and $S_d(\delta)$ corresponding to the voltage split lower band. The two figures differ only in the Hubbard interaction: in Fig.~\ref{fig:rys1} we have taken $U=5$, while for Fig.~\ref{fig:rys2} we have chosen a much larger value, namely $U=20$.

Another important, and well-visible effect of correlated hopping is the continuous shift of the voltage related minima inside the lower Hubbard bands of $G_d(\delta)$ towards negative $\delta$ values. The changes of the thermopower in this range are even more pronounced as for some range of negative $\delta$'s, $S_d$ changes sign from negative to positive with increasing $x$; c.f.\ Figs.~\ref{fig:rys1} and \ref{fig:rys2}.

In the strongly non-linear regime, a large voltage bias $eV$ splits the lower and upper Hubbard bands. However, the correlated hopping strongly affects the lower one, as is visible in Figs.~\ref{fig:rys1} and \ref{fig:rys2}, leaving the upper Hubbard band essentially unchanged. This suggests that the degree of asymmetry of the conductance and the thermopower between the two Hubbard subbands in the strongly non-linear regime provides a direct indication of the correlated hopping.

The thermoelectric power provides, at least in principle, different information about the studied system.
Roughly speaking, this is because the conductance can be argued to depend mainly on the magnitude of the density of states at the Fermi energy, while the thermopower is related to its slope. This follows from the fact that both transport coefficients are related by the celebrated Mott-Cutler formula~\cite{cutler1969}, which states that considered as a function of the chemical potential $\mu$ (or gate voltage $\delta$) the thermopower roughly follows the derivative of the conductance with respect to $\mu$. This is also approximately valid for the data presented in Fig.~\ref{fig:rys2}. 
These results demonstrate that the thermopower, being an asymmetric function of $\delta$ for $x=0$, is modified by correlated hopping ($x\ne 0$) differently for positive and negative values of $\delta$. This provides an additional clear signature of the effect of correlated hopping.

In Table~\ref{tab:table1} we summarize the symmetries of both differential transport coefficients for a quantum dot symmetrically coupled to the external electrodes.
\begin{table}[h]
	\caption{Symmetries of the differential conductance and the differential thermopower with respect to the departure from the particle-hole symmetry point ($\delta$), bias voltage ($V$), and correlated-hopping parameter ($x$), for a symmetrically coupled ($\Gamma^L=\Gamma^R$) quantum dot. }
	\centering
		\begin{tabular}{|c|c|c|}
		\hline
			 $x=0$, all $V$  &  $G_d(\delta)=G_d(-\delta)$  &  $S_d(\delta)=-S_d(-\delta)$ \\
	all $x,\delta$ &  $G_d(V)=G_d(-V)$ & $S_d(V)=S_d(-V)$ \\
	all $V,\delta$ &  $G_d(x)=G_d(2-x)$ & $S_d(x)=S_d(2-x)$ \\
		\hline
		\end{tabular}
	\label{tab:table1}
\end{table}

For non-symmetric couplings, $\Gamma^L\ne\Gamma^R$, the only remaining symmetries are $G_d(\delta,V)=G_d(-\delta,-V)$ and  $S_d(\delta,V)=-S_d(-\delta,-V)$; but these are only valid for the standard Hubbard interaction, i.e., for $x=0$. 
No other exact symmetries are permissible.

\section{Summary and conclusions}
\label{sec:sum-concl}
Non-linear transport measurements of nanostructures are known to provide interesting novel information~\cite{muralidharan2012,szukiewicz2015} about the microscopic details of the studied system. In this work, we have analysed the transport properties of a two-terminal strongly interacting quantum dot in the non-linear regime. Besides the standard Hubbard type on-site interaction between electrons on a dot, we considered another interaction term which has the structure of occupation-dependent hopping, and hence is called correlated hopping. 

Focusing, in particular, on the differential conductance and the differential Seebeck coefficient, we have shown that their line shapes as functions of gate voltage (quantified by $\delta$) uniquely indicate the existence of this important interaction term. The modifications in the line shapes due to asymmetric couplings to the external electrodes, on the other hand, are vastly different, and hence may be easily distinguished from the former. While asymmetric couplings may lead to rectification effects~\cite{kiw2023}, i.e., a non-standard dependence of the non-linear currents on the bias direction, e.g., $I(V) \ne -I(-V)$,
the correlated hopping results in conductance and thermopower line shapes that are insensitive to the bias polarisation. This provides an additional knob which can be easily used to distinguish the two sources of asymmetry, and hence to pinpoint uniquely the correlated hopping.

\acknowledgments{The work reported here has been supported in part by the National Science Center (Weave programme) through grant no.~2022/04/Y/ST3/00061.}

\end{document}